\documentclass[aps,prb,twocolumn,showpacs]{revtex4}

\usepackage{amsmath}
\usepackage{graphicx}
\usepackage{setspace}
\usepackage{bm}
\usepackage{dcolumn}
\usepackage{ulem}

\begin{document}

\title{Vortex penetration and flux relaxation with arbitrary initial conditions in non-ideal and ideal superconductors}

\author{Rongchao Ma} 
\affiliation{Edmonton, AB, Canada}
\email{marongchao@yahoo.com}

\date{\today}

\begin{abstract}
Vortex penetration and flux relaxation phenomenon carry the information about the pinning ability, and consequently current-carrying ability, of a type-II superconductor. However, the theoretical descriptions to these phenomena are currently limited to the cases with special initial conditions. A generalization to the recently developed infinite series models is presented here. It is shown that one can convert a vortex penetration process with a non-zero initial internal field into a process with a zero initial internal field by introducing some time parameters. Similarly, one can also convert a flux relaxation process starting with an arbitrary internal field into a process starting with a melting internal field by introducing a virtual time interval. Therefore, one can predict the melting internal field (or critical current density) from a flux relaxation process starting with a lower internal field. Finally, it is shown that the vortex penetration process in an ideal superconductor is strongly time dependent because of the surface barrier and internal field repulsive force. But the flux relaxation process does not occur in the ideal superconductor.
\end{abstract}


\maketitle

\section{Introduction}

In vortex penetration process, the vortices are pushed into a type-II superconductor by external driving force \cite{Sela,Hernandez,Baelus1,Berdiyorov,Baelus2,Erdin,Elistratov,Wang,Mawatari,Pissas,Ma1}. The internal field is then an increasing function of time. At lower applied fields or lower temperatures, the vortex penetration can be regarded as a process of vortices hopping between adjacent pinning centers \cite{Ma1} which can be described by the Arrhenius relation. By proposing an internal field dependent activation energy, one can obtain the corresponding time evolution equation of the internal field.\cite{Ma1} In application, a superconducting device may be repeatedly loaded and unloaded, and part of the internal field remains in the bulk of the superconductor because of pinning.\cite{Tinkham} In this case, the vortex penetration process starts with a nonzero initial internal field. However, the existing theoretical models \cite{Ma1} are only applicable to the vortex penetration process with zero initial internal field. Thus, we need to extend these theoretical models and make them applicable to the vortex penetration processes with a nonzero initial internal field.

On the other hand, in the flux relaxation process, the internal field (or current density) of a superconductor decays due to the spontaneously vortex hopping between adjacent pinning centers.\cite{Landau,Koren,Nicodemi,Hoekstra,Ma2} The internal field (or current density) is then a decreasing function of time. Using the time evolution equation of the internal field, one can determine the melting internal field from a flux relaxation measurement by applying an external field above the melting value. But this measurement may be technically difficult at lower temperatures where the melting internal field is large.\cite{Larbalestier} Therefore, we need to find a general formula which can predict the melting internal field from a flux relaxation process starting with a lower initial internal field.

Furthermore, it is known that ideal superconductors (clean superconductors) \cite{Anderson1} have a perfect crystal lattice and are free of pinning centers. This indicates that magnetic vortices cannot be pinned down in an ideal superconductor. But it is not clear how fast the vortices can penetrate into or hop out of the ideal superconductor. Is the time evolution equation of an internal field logarithmic or non-logarithmic? Thus, it is desirable to develop a mathematical model to describe the time dependence of an internal field in the vortex penetration and flux relaxation process of an ideal superconductor.

In this article we considered the fact that, in the flux relaxation and vortex penetration process, the pinning potential of the vortex does not change but the activation energy does.\cite{Ma3} Therefore, we constructed the general equations for time dependence of activation energy and of the internal field by introducing some equivalence time parameters related to the pinning potential. Using these equations, one can predict the maximum internal field in a vortex penetration process and predict the melting internal field from a flux relaxation process with an arbitrary initial internal field. We also showed that the vortex penetration into an ideal superconductor is strongly time dependent.

Because the activation energy is a function of coherence length and penetration depth, it includes information about the anisotropy of a material. Therefore, we intend to construct the theoretical models which can be applied to both high-$T_c$ and low-$T_c$ superconductors. In case that the anisotropy of the material has to be specified, we discuss it separately (Section IV). Also, in the derivation we only considered bulk pinning potential and Bean-Livingston surface barrier, but ignored geometry barriers.

\section{Vortex penetration process with arbitrary initial conditions}

The purpose of this section is to find a general expression for the time dependence of an internal field in a vortex penetration process with an arbitrary initial internal field. To do this, we need the internal field dependence of activation energy and time dependence of activation energy.

\subsection{Internal field dependence of activation energy in vortex penetration}

The Bean model \cite{Bean2} shows that, under an applied field $B_a$, the internal field should approach a maximum value $B_e$ when the superconductor reaches an equilibrium state. Because of the surface screening effect (or Meissner effect), $B_e$ must be smaller than $B_a$, that is, $B_e < B_a$.

\paragraph{Series expression of activation energy}

In vortex penetration process, the repulsive force of the internal field $B$ prevents vortex motion and reduces vortex hopping rate. The activation energy $U_p$ is then an increasing function of $B$. Thus, $dU_p/dB>0$. In early study \cite{Ma1}, $U_p$ is expressed as a series of $B$ (or magnetization $M$). For the convenience of calculating the equilibrium field $B_e$, here I expand $U_p$ as a series of the normalized field $\sigma=B/B_e$, that is, 
\begin{equation}
\label{UpSeries}
U_p(\sigma) = U_{p0} + \sum\limits_{l=1}^\infty a_l \sigma^l,
\end{equation}
where $U_{p0}=U_{BL}+U_c$ is the activation energy at vanishing internal field, $U_{BL}$ is the Bean-Livingston surface barrier, \cite{Bean1} and $U_c$ is pinning potential.

\paragraph{Activation energy at initial internal field}

If the initial internal field $B_i>0$ (at $t=0$), then the corresponding initial activation energy $U_i \geq U_{p0}$. Using Eq.(\ref{UpSeries}), we can express $U_i$ in terms of $\sigma_i=B_i/B_e$, that is,
\begin{equation}
\label{UiBi}
U_i =  U_{p0} + \sum\limits_{l=1}^\infty a_l \sigma_i^l.
\end{equation}

\paragraph{Activation energy at equilibrium internal field}

At the equilibrium field $B_e$, the corresponding equilibrium activation energy is
\begin{equation}
\label{Ue}
U_e = U_{p0} + \sum\limits_{l=1}^\infty a_l.
\end{equation}

On the basis of the above discussion, we can now draw a schematic diagram of $U_p(B)$, the internal field dependence of activation energy in a vortex penetration process, as shown in Fig. 1(a).

\begin{figure}[htb]
\label{Fig1}
\begin{center}
\includegraphics[height=0.50\textwidth]{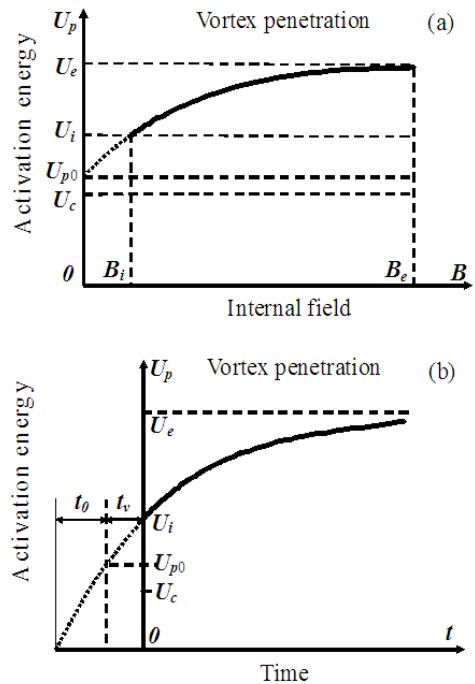}
\caption{Schematic diagram of $U_p$, the activation energy of a vortex penetration process. $U_c$ is pinning potential, $U_{p0}$ is the activation energy at vanishing internal field, $U_i$ is initial activation energy, $U_e$ is equilibrium activation energy, $B_i$ is initial internal field and $B_e$ is equilibrium internal field. (a) Internal field dependence of activation energy. $U_p$ is an increasing function of internal field $B$. (b) Time dependence of activation energy. $U_p$ is an increasing function of time $t$. }
\end{center}
\end{figure}

Inverting Eq.(\ref{UpSeries}) and using definition $\sigma=B/B_e$, we obtain the expression of internal field $B$ in terms of the activation energy $U_p$, that is,
\begin{equation}
\label{BUp}
B(U_p) = B_e \sum\limits_{l=1}^\infty b_l (U_p - U_{p0})^l,
\end{equation}
where the coefficients $b_l$ are \cite{Ma1}
\begin{equation}
\label{bns}
\begin{aligned}
b_l =&\frac{1}{a_1^l} \frac{1}{l} \sum\limits_{s,t,u \cdots} (-1)^{s+t+u+\cdots} \cdot \\
     &\frac{l(l+1)\cdots(l-1+s+t+u+\cdots)}{s!t!u!\cdots} \cdot \\
     &\left(\frac{a_2}{a_1}\right)^s \left(\frac{a_3}{a_1}\right)^t \left(\frac{a_4}{a_1}\right)^u \cdots, \\    
\end{aligned}
\end{equation}
and $s+2t+3u+\cdots=l-1$. On considering the symmetry between Eq.(\ref{UpSeries}) and Eq.(\ref{BUp}), we can obtain the inverse coefficients $a_l$ by doing a commutation to the coefficients $b_l \leftrightarrow a_l$.

Eq.(\ref{BUp}) shows that, to obtain the time dependence of internal field $B(t)$, we still need the time dependence of activation energy $U_p(t)$. Let us discuss this in the next section.

\subsection{Time dependence of activation energy in vortex penetration}

\subsubsection{Solution of the time dependence of activation energy}

We have shown that the activation energy of the vortex penetration process, $U_p$, is an increasing function of the internal field $B$. Because $B$ is an increasing function time $t$ ($dB/dt>0$), $U_p$ is an increasing function of $t$. Using the Arrhenius relation, we have \cite{Ma1}
\begin{equation}
\label{UpDiff}
\frac{dB}{dt} = C e^{-U_p/kT},
\end{equation} 
where $C$ is a positive proportional constant, $k$ is the Boltzmann constant and $T$ is temperature.\cite{Tinkham} 

Consider a superconductor with an initial internal field $B_i$ and initial activation energy $U_i$. Now apply an external field $B_a$ to the superconductor at zero time ($t=0$). As the time increases to $t$, the activation energy increases from $U_i$ to $U_p$. Rewrite Eq.(\ref{UpDiff}) and integrate it on both sides
\[ \int_{U_i}^{U_p} e^{U_p/kT}dU_p = \int_0^t C\frac{dU_p}{dB} dt.  \]
With logarithmic accuracy, we obtain the following equation \cite{Ma1}
\begin{equation}
\label{UpTime1}
U_p(t) = kT ln\left( e^{U_i/kT} + \frac{t}{\tau} \right),
\end{equation} 
where $\tau = kT/[C(dU_p/dB)]$ is a short time scale parameter \cite{Ma1}. Eq.(\ref{UpTime1}) describes the general time dependence of activation energy in a vortex penetration process, $U_p(t)$, which is an increasing function of time $t$.

\subsubsection{Meaning of virtual time intervals}

To assist physical understanding and easy calculation, let us define the following time parameters
\begin{subequations}
\begin{eqnarray}
t_i &=& \tau \left( e^{U_i/kT}-1 \right),  \label{tiU} \\
t_0 &=& \tau \left( e^{U_{p0}/kT}-1 \right),  \label{t0U} \\
t_v &=& t_i-t_0 = \tau \left( e^{U_i/kT}-e^{U_{p0}/kT} \right),  \label{tvU}
\end{eqnarray}
\end{subequations}
where $U_{p0}$ is the activation energy at vanishing internal field (Eq.(\ref{UpSeries})) and $U_i$ is the initial activation energy (Eq.(\ref{UiBi})). Thus, Eq.(\ref{UpTime1}) becomes
\begin{equation}
\label{UpTime2} 
U_p(t) = kT ln\left(1 + \frac{t_i+t}{\tau} \right) = kT ln\left( 1 + \frac{t_0 + t_v + t}{\tau} \right).
\end{equation}

\paragraph{Meaning of $t_v$}

In Eq.(\ref{UpTime2}), if we let $t \rightarrow 0$, then $U_p \rightarrow U_i$ (use Eq.(\ref{t0U}) and Eq.(\ref{tvU})). If we further let $t_v \rightarrow 0$, then $U_i \rightarrow U_{p0}$. This indicates that the physical meaning of $t_v$ is: \textit{the virtual time interval during which the activation energy $U_p(t)$ increase from $U_{p0}$ to $U_i$. }

Putting $t_v = 0$ in Eq.(\ref{UpTime2}), we have
\begin{equation}
\label{UpTime3} 
U_p(t) = kT ln\left(1 + \frac{t_0+t}{\tau} \right).
\end{equation} 

Eq.(\ref{tvU}) shows that, if $t_v=0$, then $U_i=U_{p0}$. From Eq.(\ref{UiBi}), we have the initial internal field $B_i=0$. Thus, Eq.(\ref{UpTime3}) is the time dependence of activation energy in a vortex penetration process with zero initial internal field.\cite{Ma2}

By doing a time transformation $t'=t_v+t$, we can convert Eq.(\ref{UpTime2}) into the form of Eq.(\ref{UpTime3}). In other words, we can convert a vortex penetration process with nonzero initial internal field into a process with zero initial internal field by introducing a time parameter $t_v$ as defined in Eq.(\ref{tvU}). This can be clearly seen in the later discussion on the time dependence of internal field.

\paragraph{Meaning of $t_0$}

Because $U_{p0}$ is the activation energy at vanishing internal field (Eq.(\ref{UpSeries})), a superconductor with $U_{p0}=0$ means that it is free of pinning. The superconductor is then an ideal superconductor, or clean superconductors \cite{Anderson1} (Here I assume that the superconductor is isotropic. The layered superconductor is discussed in section IV). Thus, using Eq.(\ref{t0U}) we can conclude that the physical meaning of $t_0$ is: \textit{the time parameter equivalent to the potential $U_{p0}$ in a non-ideal superconductor}.

Putting $t_0=0$ in Eq.(\ref{UpTime3}), we have
\begin{equation}
\label{UpTime4} 
U_p(t) = kT ln\left(1 + \frac{t}{\tau} \right).
\end{equation}

$t_v=0$ means that the initial internal field $B_i=0$ (Eq.(\ref{UpTime3})), and $t_0=0$ means that the superconductor is an ideal superconductor (Eq.(\ref{t0U})). Thus, Eq.(\ref{UpTime4}) is the time dependence of the activation energy of a vortex penetration process in an ideal superconductor with zero initial internal field.

By doing a time transformation $t''=t_0+t$, we can convert Eq.(\ref{UpTime3}) into the form of Eq.(\ref{UpTime4}). In other words, we can convert a vortex penetration process of a nonideal superconductor into a process of an ideal superconductor by introducing a time parameter $t_0$, as defined in Eq.(\ref{t0U}).

From the above discussions we know that there is a equivalence relation between the time $t$ and activation energy $U_p$. To see this, let us invert Eq.(\ref{UpTime4}),
\begin{equation}
\label{tU} 
t = \tau \left( e^{U_p/kT}-1 \right).
\end{equation}

Eq.(\ref{tU}) shows that, in an ideal superconductor with zero initial internal field, the activation energy of a vortex penetration process can be mapped into a time parameter, that is, $f:U_p \rightarrow t$, $U_p \in [0,U_e]$. Although Eq.(\ref{tU}) is obtained from an ideal superconductor, it can generate all the relations for nonideal superconductors because the potential $U_{p0}$ is equivalent to a time parameter $t_0$.

The time dependence of activation energy in a vortex penetration process is schematically shown in Fig. 1(b).

\subsection{Time dependence of internal field in vortex penetration}

The time dependence of internal field $B(t)$ can be obtained by combing the internal field dependence of activation energy $U_p(\sigma)$ with the time dependence of activation energy $U_p(t)$.

\subsubsection{Infinite series activation energy}

Substituting Eq.(\ref{UpTime2}) into Eq.(\ref{BUp}), we have
\begin{equation}
\label{Bwp}
B(t) = B_e \sum\limits_{l=1}^\infty b_l [w_p(t)]^l,
\end{equation}
where
\begin{equation}
\label{wpt}
w_p(t) = kT ln\left(1 + \frac{t_0+t_v+t}{\tau} \right) - U_{p0}.
\end{equation}

Eq.(\ref{Bwp}) describes the time dependence of the internal field $B(t)$. Fitting Eq.(\ref{Bwp}) to the experimental data, we can obtain various parameters and then calculate the activation energy.\cite{Ma1}

It should be mentioned that, to determine $B_e$, we need both Eq.(\ref{Bwp}) and the inverse of Eq.(\ref{Ue}),
\begin{equation}
\label{BlConstraint}
1 = \sum\limits_{l=1}^\infty b_l (U_e - U_{p0})^l.
\end{equation}
This can be obtained by simply putting $B=B_e$ and $U_p=U_e$ in Eq.(\ref{BUp}). 

If we are only interested in finding out the time dependence of the internal field, then we can simplify Eq.(\ref{Bwp}) by absorbing $B_e$ into the coefficients $b_l$. Putting $h_l= b_l B_e$ in Eq.(\ref{Bwp}), we have
\begin{equation}
\label{Bwp2}
B(t) = \sum\limits_{l=1}^\infty h_l [w_p(t)]^l.
\end{equation}
This is equal to expanding the activation energy as a function of the internal field $B$ (Ref.~\onlinecite{Ma1}), that is, $U_p(B) = U_{p0} + \sum_{l=1}^\infty a_l B^l$.

Putting $t=0$ in Eq.(\ref{Bwp}), we obtain the initial internal field
\begin{equation}
\label{Bi}
B_i = B_e \sum\limits_{l=1}^\infty b_l \left[kT ln\left(1 + \frac{t_0+t_v}{\tau} \right) - U_{p0}\right]^l.
\end{equation}

Putting $t_v=0$ in Eq.(\ref{Bi}) and using Eq.(\ref{t0U}), we have $B_i=0$. This is consistent with the discussion on Eq.(\ref{UpTime3}), in which we interpreted the physical meaning of $t_v$ using the activation energy, that is, the time interval during which the activation energy increase from $U_{p0}$ to the initial value $U_i$. Now we can explain it in terms of internal field: \textit{$t_v$ is the time interval during which the internal field increases from $0$ to the initial value $B_i$}.

Eq.(\ref{Bwp}) also shows that a vortex penetration process with an initial internal field $B_i$ can be converted into a process with zero initial internal field by introducing a virtual time interval $t_v$.

In Eq.(\ref{tiU}) and Eq.(\ref{tvU}) we express the time parameters $t_i$ and $t_v$ in terms of the initial activation energy $U_i$. But in practice the measurable physical quantity is the initial internal field $B_i$. Therefore, it should be more convenient to express $t_i$ and $t_v$ in terms of $B_i$ ($\sigma_i=B_i/B_e$). To do this, a direct way is to invert Eq.(\ref{Bi}), but an easier way is to use Eq.(\ref{UiBi}).

Substituting Eq.(\ref{UiBi}) into Eq.(\ref{tiU}), we have 
\begin{equation}
\label{tiBi}
t_i = \tau \left[ e^{U_{p0}/kT} \prod\limits_{l=1}^\infty e^{a_l \sigma_i^l/kT} -1 \right].
\end{equation}

Substituting Eq.(\ref{UiBi}) into Eq.(\ref{tvU}), we have 
\begin{equation}
\label{tvBi}
t_v = \tau e^{U_{p0}/kT} \left[ \prod\limits_{l=1}^\infty e^{a_l \sigma_i^l/kT} -1 \right].
\end{equation}

If we further put $t_0=0$ in Eq.(\ref{Bi}), we still have $B_i=0$. Therefore, in a nonideal superconductor, the physical meaning of $t_0$ cannot be explained in terms of an internal field.

\subsubsection{Quadratic activation energy}

Without inelastic deformation \cite{Ma1}, a noninteracting elastic vortex system has $a_l=0$ ($l>2$). Thus, Eq.(\ref{UpSeries}) reduces to the following quadratic activation energy
\begin{equation}
\label{UpQuadratic}
U_p(B) = U_{p0} + a_1 \sigma + a_2 \sigma^2,
\end{equation}
and Eq.(\ref{Ue}) reduces to $U_e=U_{p0} + a_1 + a_2$.

Substituting Eq.(\ref{UpTime2}) into Eq.(\ref{UpQuadratic}), we have (choose one of the solutions that is an increasing function of time)
\begin{equation}
\label{BQuadratic}
B(t) = B_e \frac{a_1}{2 a_2} \left[ \sqrt{1 + 4\frac{a_2}{a_1^2} w_p(t) } - 1 \right],
\end{equation}
where $w_p(t)$ is defined by Eq.(\ref{wpt}).

Putting $t=0$ in Eq.(\ref{BQuadratic}), we have
\begin{equation}
\label{BiQuadratic}
B_i = B_e \frac{a_1}{2 a_2} \left[ \sqrt{1 + 4\frac{a_2}{a_1^2} \left[kT ln\left(1 + \frac{t_0+t_v}{\tau} \right) - U_{p0}\right] } - 1 \right].
\end{equation}

Eq.(\ref{tiBi}) reduces to
\begin{equation}
\label{tiBiQuadratic}
t_i = \tau \left[ e^{(U_{p0}+ a_1 \sigma_i + a_2 \sigma_i^2)/kT} -1 \right],
\end{equation}
and Eq.(\ref{tvBi}) reduces to
\begin{equation}
\label{tvBiQuadratic}
t_v = \tau e^{U_{p0}/kT} \left[e^{(a_1 \sigma_i + a_2 \sigma_i^2)/kT} -1 \right].
\end{equation}

\subsubsection{Linear activation energy}

In a noninteracting rigid vortex system, the elastic deformation vanishes \cite{Ma1}. Thus, the coefficient $a_2=0$ and Eq.(\ref{UpSeries}) reduces to the following linear activation energy
\begin{equation}
\label{UpLinear}
U_p(B) = U_{p0} + a_1 \sigma,
\end{equation}
and Eq.(\ref{Ue}) reduces to $U_e=U_{p0} + a_1$.

Substituting Eq.(\ref{UpTime2}) into Eq.(\ref{UpLinear}), we have 
\begin{equation}
\label{BLinear}
B(t) = \frac{B_e}{a_1} w_p(t).
\end{equation}
where $w_p(t)$ is defined by Eq.(\ref{wpt}).

Putting $t=0$ in Eq.(\ref{BLinear}), we have
\begin{equation}
\label{BiLinear}
B_i = \frac{B_e}{a_1} \left[kT ln\left(1 + \frac{t_0+t_v}{\tau} \right) - U_{p0}\right].
\end{equation}

Eq.(\ref{tiBi}) reduces to
\begin{equation}
\label{tiBiLinear}
t_i = \tau \left[ e^{(U_{p0}+a_1 \sigma_i)/kT} -1 \right],
\end{equation}
and Eq.(\ref{tvBi}) reduces to
\begin{equation}
\label{tvBiLinear}
t_v = \tau e^{U_{p0}/kT} \left(e^{a_1 \sigma_i/kT} -1 \right).
\end{equation}

\section{Flux relaxation process with arbitrary initial conditions}

In the study of flux relaxation, a vortex's activation energy is usually expressed as a function of current density and the corresponding time dependence of current density can be derived.\cite{Feigel'man2,Feigel'man1,Zeldov1,Zeldov2,Anderson2, Ma2} In the present work, we are going to discuss the flux relaxation process in an ideal superconductor. Because the flux relaxation process is essentially a process of vortex motion, it should make more sense to use internal field instead of current density. In case of need, one can convert the formulas in terms of internal field into formulas in terms of current density or magnetization by doing a replacement.

\subsection{Internal field dependence of activation energy in flux relaxation}

At a temperature $T$, a vortex system will melt as the internal field increase to a value $B_m$. Corresponding to this $B_m$, there should be a melting field as defined in the conventional magnetic phase of type-II superconductors. The melting field is not an internal field, but an external field.

It should be emphasized that the melting internal field $B_m$ is different from the equilibrium field $B_e$ in the vortex penetration process as discussed in Eq.(\ref{UpSeries}). In the flux relaxation process, $B_m$ is the maximum internal field for the vortex system to keep a lattice structure \cite{Tinkham} (corresponding to the external melting field), which is determined by the temperature, pinning and material of the superconductor. But in a vortex penetration process, $B_e$ is the maximum magnetic field can penetrate into the superconductor under an applied magnetic field $B_a$, which is determined by the applied field $B_a$.

\paragraph{Series expression of activation energy}

In the flux relaxation process, the repulsive force between vortices can enhance the vortex motion and increases vortex hopping rate. The activation energy $U_r$ is then a decreasing function of the internal field $B$. This indicates that $dU_r/dB<0$. According to an early study \cite{Ma2}, we can express $U_r$ as a series of internal field $B$. For the convenience of calculating the melting internal field $B_m$, I expand $U_r$ as a series of the normalized internal field $\lambda=B/B_m$, that is,
\begin{equation}
\label{UrSeries}
U_r(\lambda) = U_{r0} - \sum\limits_{l=1}^\infty c_l \lambda^l,
\end{equation}
where $U_{r0}=U_c-U_{im}$. The parameter $U_c$ is the pinning potential inside the bulk and $U_{im}=(\Phi_0/4 \pi \lambda)^2 K_0(2x/\lambda)$ is a reduction to the activation energy caused by the surface imaging force.\cite{Ma2,Bean1} Eq.(\ref{UrSeries}) shows that $U_{r0}$ is the activation energy of vortices at vanishing driving force ($B=0$) and is also the maximum activation energy of the entire flux relaxation process.

\paragraph{Activation energy at initial internal field}

Let $B_i$ be the initial internal field at time $t=0$ and $U_i$ be the corresponding initial activation energy. Putting $B=B_i$ (or $\lambda_i=B_i/B_m$) in Eq.(\ref{UrSeries}), we have
\begin{equation}
\label{Uiji}
U_i = U_{r0} - \sum\limits_{l=1}^\infty c_l \lambda_i^l.
\end{equation}

Since the activation energy $U_r$ is a decreasing function of the internal field, Eq.(\ref{Uiji}) indicates that $U_i \leq U_{r0}$.

\paragraph{Activation energy at melting internal field}

At the melting internal field $B_m$, the activation energy $U_r$ is zero. Eq.(\ref{UrSeries}) gives
\begin{equation}
\label{Ur0}
U_{r0} = \sum\limits_{l=1}^\infty c_l.
\end{equation}
This is a constraint condition for the coefficients $c_l$.

On the basis of the above discussions, we can now draw a schematic diagram of $U_r(B)$, the field dependence of activation energy in a flux relaxation process, as shown in Fig. 2(a).

\begin{figure}[htb]
\label{Fig2}
\begin{center}
\includegraphics[height=0.50\textwidth]{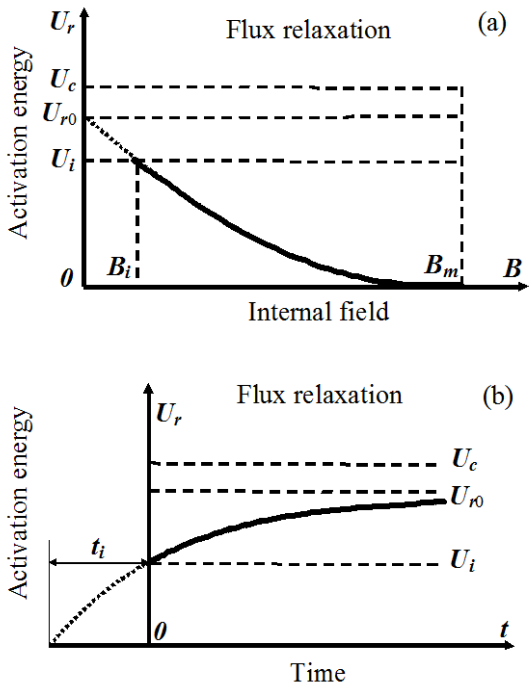}
\caption{Schematic diagram of $U_r$, the activation energy of a flux relaxation process. $U_c$ is pinning potential, $U_{r0}$ is the activation energy at vanishing internal field, $U_i$ is initial activation energy, $B_i$ is initial internal field and $B_m$ is melting internal field. (a) Internal field dependence of activation energy. $U_r$ is a decreasing function of internal field $B$. (b) Time dependence of activation energy. $U_r$ is an increasing function of time $t$. }
\end{center}
\end{figure}

Inverting Eq.(\ref{UrSeries}) and using the definition $\lambda=B/B_m$, we obtain the expression of internal field $B$ in terms of activation energy $U_r$, that is,
\begin{equation}
\label{jUr}
B(U_r) = B_m \sum\limits_{l=1}^\infty d_l (U_{r0} - U_r)^l,
\end{equation}
where the coefficients $d_l$ and $c_l$ has the same relation as that of $b_l$ and $a_l$ (See Eq.(\ref{bns})).

Eq.(\ref{jUr}) shows that we can obtain the time dependence of the internal field $B(t)$ by finding out the time dependence of the activation energy $U_r(t)$. This is further studied in the next section.

\subsection{Time dependence of activation energy in flux relaxation}

In the flux relaxation process, the vortex's activation energy $U_r$ is a decreasing function of the internal field $B$. Because $B$ is a decreasing function of time $t$ ($dB/dt<0$), $U_r$ is an increasing function of $t$. Similar to Eq.(\ref{UpDiff}), we have
\begin{equation}
\label{UrDiff}
\frac{dB}{dt} = - C e^{-U_r/kT}
\end{equation} 
where $C$ is a positive proportional constant.

Consider a superconductor in which an arbitrary internal field starts to decay at zero time ($t=0$). The initial activation energy of the vortices is $U_i$. As time increases to $t$, the activation energy increases to $U_r$. Similar to Eq.(\ref{UpTime2}), we obtain the following equation \cite{Ma3}
\begin{equation}
\label{UrTime}
U_r(t) = kT ln\left(1 + \frac{t_i+t}{\tau} \right),
\end{equation}
where
\begin{equation}
\label{tiUr}
t_i = \tau \left( e^{U_i/kT}-1 \right)
\end{equation}
and $\tau = - kT/[C(dU_r/dB)]$. Here we include a negative sign to ensure that $\tau$ is a positive number because in the flux relaxation process the activation energy $U_r$ is a decreasing function of the internal field $B$, i.e., $dU_r/dB<0$.

Eq.(\ref{UrTime}) shows that the activation energy of a flux relaxation process starting with an internal field below the melting internal field can be converted into that of a process starting with the melting internal field by introducing a virtual time interval $t_i$. The physical meaning of $t_i$ is: \textit{the virtual time interval during which the activation energy $U_r(t)$ increases from $0$ to the initial value $U_i$}.

It should be mentioned that the time parameter $t_i$ in Eq.(\ref{UrTime}) can not be split into $t_0$ and $t_v$ as we did in Eq.(\ref{UpTime2}). This is because in the flux relaxation process, the activation energy at vanishing internal field, $U_{r0}$, is the maximum activation energy. The initial activation energy $U_i$ satisfies $U_i \leq U_{r0}$. Thus, $U_i$ cannot be further split. But in vortex penetration process the activation energy at vanishing internal field, $U_{p0}$, is the minimum activation energy. The initial activation energy $U_i$ satisfies $U_i \geq U_{p0}$. Thus, $U_i$ can be split into two parts.

Eq.(\ref{UrTime}) describes the general time dependence of the activation energy in a flux relaxation process, $U_r(t)$, which is an increasing function of time $t$. It turns out that both the activation energy of the flux relaxation process $U_r(t)$ (Eq.(\ref{UrTime})) and the activation energy of vortex penetration process $U_p(t)$ (Eq.(\ref{UpTime2}))are increasing function of time $t$.

The time dependence of activation energy in a flux relaxation process is schematically shown in Fig. 2(b).

\subsection{Time dependence of internal field in flux relaxation}

The time dependence of the internal field $B(t)$ can be obtained by combing the time dependence of activation energy $U_r(t)$ (Eq.(\ref{UrTime})) with a detailed internal field dependence of activation energy $U_r(B)$.

\subsubsection{Infinite series activation energy}

The infinite series activation energy \cite{Ma2} is shown in Eq.(\ref{UrSeries}) and its inverse is shown in Eq.(\ref{jUr}). Substituting Eq.(\ref{UrTime}) into Eq.(\ref{jUr}), we obtain the time dependence of the internal field
\begin{equation}
\label{jwr}
\begin{aligned}
B(t) = B_m \sum\limits_{l=1}^\infty d_l [w_r(t)]^l,
\end{aligned}
\end{equation}
where
\begin{equation}
\label{wrt}
w_r(t) = U_{r0} - kTln\left(1+\frac{t_i+t}{\tau}\right).
\end{equation}

Putting $t=0$ in Eq.(\ref{jwr}), we obtain the initial internal field
\begin{equation}
\label{ji}
B_i = B_m \sum\limits_{l=1}^\infty d_l \left[U_{r0} - kTln\left(1+\frac{t_i}{\tau}\right)\right]^l.
\end{equation}

Putting $t_i = 0$ in Eq.(\ref{tiUr}), we have $U_i=0$. It means that the flux relaxation process starts with the melting internal field, i.e., $B_i = B_m$. Thus, Eq.(\ref{ji}) gives
\begin{equation}
\label{blConstraint}
\sum_{l=1}^\infty d_l U_{r0}^l = 1.
\end{equation}

Eq.(\ref{blConstraint}) is a constraint condition of the coefficients $d_l$. It is also the inverse of Eq.(\ref{Ur0}) (rewriting as $ U_{r0} = \sum_{l=1}^\infty c_l \times 1 $).

Eq.(\ref{jwr}) describes the time dependence of the internal field $B(t)$. It shows that a flux relaxation process starting with an initial internal field $B_i$ can be converted into a process starting with the melting internal field $B_m$ by introducing a virtual time interval $t_i$. Fitting Eq.(\ref{jwr}) to experimental data, we can obtain the melting internal field $B_m$. This means that we can calculate $B_m$ from a flux relaxation process with an arbitrary initial internal field $B_i$.

In Eq.(\ref{tiUr}), we expressed the virtual time interval $t_i$ in terms of the initial activation energy $U_i$. Let us now express $t_i$ in terms of the initial internal field $B_i$ ($\lambda_i=B_i/B_m$). Substituting Eq.(\ref{Uiji}) into Eq.(\ref{tiUr}), we have
\begin{equation}
\label{tiji}
t_i = \tau \left[ e^{U_{r0}/kT}\prod\limits_{l=1}^\infty e^{-c_l \lambda_i^l/kT} - 1 \right].
\end{equation}

When deriving Eq.(\ref{UrTime}), we explained the physical meaning of the virtual time interval $t_i$ using activation energy. Now, we can explain it using the internal field from Eq.(\ref{jwr}): \textit{$t_i$ is the time interval during which the internal field reduces from the melting value $B_m$ to the initial value $B_i$}.

\subsubsection{Quadratic activation energy} 

In a noninteracting elastic vortex system with vanishing inelastic deformation \cite{Ma2}, the coefficient $c_l=0$ ($l>2$). Thus, Eq.(\ref{UrSeries}) reduces to the following quadratic activation energy
\begin{equation}
\label{UrQuadratic}
U_r(\lambda) = U_{r0} - c_1 \lambda - c_2 \lambda^2,
\end{equation}
where $c_1 + c_2 = U_{r0}$ (Using Eq.(\ref{Ur0})).

Substituting Eq.(\ref{UrTime}) into Eq.(\ref{UrQuadratic}), we obtain the time dependence of internal field $B(t)$ (choose one of the solutions which is a decreasing function of $t$)
\begin{equation}
\label{jQuadratic}
B(t) = B_m \frac{c_1}{2 c_2} \left[ \sqrt{1 + 4\frac{c_2}{c_1^2} w_r(t) }-1 \right],
\end{equation}
where the function $w_r(t)$ is defined in Eq.(\ref{wrt}).

Putting $t=0$ in Eq.(\ref{jQuadratic}), we have
\begin{equation}
\label{jiQuadratic}
B_i = B_m \frac{c_1}{2 c_2} \left[ \sqrt{1 + 4\frac{c_2}{c_1^2} \left[U_{r0} - kTln\left(1+\frac{t_i}{\tau}\right)\right] }-1 \right]
\end{equation}
and Eq.(\ref{tiji}) reduces to
\begin{equation}
\label{tiji2}
t_i = \tau \left[ e^{(U_{r0} - c_1 \lambda_i - c_2 \lambda_i^2)/kT} - 1 \right].
\end{equation}

\subsubsection{Linear activation energy}

In a noninteracting rigid vortex system \cite{Ma2}, the coefficient $c_2=0$. Eq.(\ref{UrQuadratic}) reduces to the following linear activation energy \cite{Anderson2}
\begin{equation}
\label{UrLinear}
U_r(\lambda) = U_{r0} - c_1 \lambda,
\end{equation}
where $c_1 = U_{r0}$ (Using Eq.(\ref{Ur0})).

Substituting Eq.(\ref{UrTime}) into Eq.(\ref{UrLinear}), we have 
\begin{equation}
\label{jLinear}
B(t) = \frac{B_m}{c_1} w_r(t).
\end{equation}
where the function $w_r(t)$ is defined in Eq.(\ref{wrt}). 

Putting $t=0$ in Eq.(\ref{jLinear}), we have
\begin{equation}
\label{jiLinear}
B_i = \frac{B_m}{c_1} \left[U_{r0} - kTln\left(1+\frac{t_i}{\tau}\right)\right]
\end{equation}
and Eq.(\ref{tiji}) reduces to
\begin{equation}
\label{tiji3}
t_i = \tau \left[ e^{(U_{r0} - c_1 \lambda_i)/kT} - 1 \right].
\end{equation}

\section{Discussion}

\subsection{Vortex motion in ideal superconductors}

An ideal superconductor has perfect crystal lattice and is free of defects.\cite{Anderson1} But for a layered superconductor, even if it is ``ideal'', the space between the layers can still behave as pinning centers.\cite{Tinkham} Thus, a vortex can be pinned down when it lies between two layers. In this section I am only interested in studying a physical system in which the vortices are free to move. Therefore, I assumed that the ideal superconductor under consideration is isotropic. It is not only free of pinning centers, but also free of any pinning potential, i.e., $U_c=0$.

\subsubsection{Vortex penetration}
 
In an ideal superconductor, the pinning potential is $U_c=0$. Thus, the internal field dependence of activation energy of a vortex penetration process, Eq.(\ref{UpSeries}), reduces to
\begin{equation}
\label{UpSeries2}
U_p(\sigma) = U_{BL} + \sum\limits_{l=1}^\infty c_l \sigma^l.
\end{equation}

Eq.(\ref{UpSeries2}) is normal, which indicates that the vortex penetration process may occur in an ideal superconductor. 

Let us now check the time dependence of an internal field in the ideal superconductor. First, use Eq.(\ref{UpTime1}) and rewrite Eq.(\ref{Bwp2}) as
\begin{equation}
\label{Bwp3}
B(t) = \sum\limits_{l=1}^\infty h_l \left\{ kT ln\left[e^{(U_i-U_{p0})/kT} + \frac{t}{\tau e^{U_{p0}/kT}} \right] \right\}^l.
\end{equation}

Putting $U_c=0$ in Eq.(\ref{Bwp3}), we have
\begin{equation}
\label{BWpIdeal}
B(t) = \sum\limits_{l=1}^\infty h_l \left\{ kT ln\left[e^{(U_i-U_{BL})/kT} + \frac{t}{\tau e^{U_{BL}/kT}} \right] \right\}^l.
\end{equation}

Since $(U_i-U_{BL}) \geq 0$, we have $e^{(U_i-U_{BL})/kT} \geq 1$. Thus, $B(t) \geq 0$. The simulation of Eq.(\ref{BWpIdeal}) is shown in Fig. 3. 

\begin{figure}[htb]
\label{Fig3}
\begin{center}
\includegraphics[height=0.3\textwidth]{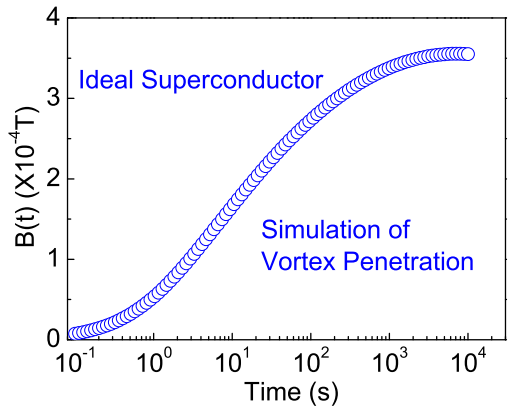}
\caption{(Color online) Simulation of vortex penetration into an ideal superconductor. The time dependence of an internal field is simulated with equation: $B_p(t) = h_1 w_p(t) + h_2 w_p^2(t)$, where $w_p(t)= (25k) \times ln(1+t)$, $h_1=8 \times 10^{-5}/(25k)$, $h_2= -4.5 \times 10^{-6}/(25k)^2$ (See Eq.(\ref{BWpIdeal})). Here I choose a position away form the surface, where $U_{BL}=0$. Also, I choose $\tau=1$ and $B_i=0$. In an ideal superconductor the pinning potential is $U_c=0$, we have $U_i=U_{BL}=0$.}
\end{center}
\end{figure}

One can see that, in an ideal superconductor, the internal field $B$ is an increasing function of time $t$. This shows that although an ideal superconductor does not have pinning centers, the vortex penetration into this superconductors is strongly time dependent. The reason is that the repulsive force of the internal field (a field gradient formed in the bulk) prevents the vortices penetrating into the bulk of the superconductor. The vortex motion is then retarded.

\subsubsection{Flux relaxation}
 
In an ideal superconductor, the pinning potential is $U_c=0$. Thus, the field dependence of the activation energy of a flux relaxation process, Eq.(\ref{UrSeries}), reduces to
\begin{equation}
\label{UrSeries2}
U_r(\lambda) = -U_{im} - \sum\limits_{l=1}^\infty c_l \lambda^l.
\end{equation}
This shows that $U_r(\lambda)<0$. However, for a flux relaxation process to occur, the activation energy must be positive. Thus, the flux relaxation process does not occur in an ideal superconductor.

Let us now check the time dependence of the internal field in an ideal superconductor. Putting $U_c=0$ in Eq.(\ref{jwr}), we have
\begin{equation}
\label{jwrIdeal}
B(t) = B_m \sum\limits_{l=1}^\infty d_l \left[ -U_{im}- kTln\left(1+\frac{t_i+t}{\tau}\right) \right]^l.
\end{equation}

Eq.(\ref{jwrIdeal}) shows that $B(t)<0$. Because $B(t)$ is nonnegative, we must have $d_l=0$ and $B(t)=0$. This means that the internal field vanishes after the external field is reduced to zero.

Thus, we can conclude that, in an ideal superconductor, the flux relaxation process does not occur. The reason is that the ideal superconductor is free of pinning centers, and vortices cannot be pinned down in the bulk of the superconductor.

\subsection{Inflection point of $B(t)-t$ curve}

An earlier study has shown that the $B(t)-t$ curves display inflection points.\cite{Ma2} The curves have concave shapes at short times and then change to convex shapes with increasing time. Here I show that this phenomenon does not occur when the initial internal field $B_i$ reaches a critical value.

For simplicity, let us keep the terms in Eq.(\ref{Bwp}) to the second order $B(t)=B_e [b_1w(t)+b_2w^2(t)]$. (Ref.~\onlinecite{Ma2}) The second derivative of $B(t)$ is 
\begin{equation}
\label{SecDerivation}
\frac{d^2B}{dt^2}= g(t) \left[1-\frac{b_1}{2b_2kT}-ln\left(1+\frac{t_i+t}{\tau}\right)+\frac{U_{p0}}{kT}\right],
\end{equation}
where $g(t)= 2b_2 B_e [(\tau + t_i +t)/kT]^{-2}$.

Letting $d^2B/dt^2=0$, we obtain the inflection point time 
\begin{equation}
\label{Inflection}
t^*= \tau \left[ exp \left(1 - \frac{b_1}{2b_2kT} + \frac{U_{p0}}{kT} \right) -1 \right] - t_i.
\end{equation}

Since $t^*$ is a time parameter, it must be positive. If $t_i \geq \tau \left \{ exp \left[1 - b_1/(2b_2kT) + U_{p0}/(kT) \right] -1 \right\}$, then the inflection point vanishes.

\section{Conclusion}

In vortex penetration and flux relaxation process, vortex's activation energy increases with increasing time, but its pinning potential does not change. Using this idea, one can convert a vortex penetration process with a nonzero initial internal field into a process with a zero initial internal field by introducing some time parameters. Similarly, one can also convert a flux relaxation process starting with a lower internal field into a process starting with a melting internal field by introducing a virtual time interval. This enables us to predict the melting internal field from a flux relaxation measurement by applying a lower magnetic field to the superconductor. Although ideal superconductors do not have pinning centers, the vortex penetration into these superconductors is still time dependent because of the repulsive force of the internal field. But the flux relaxation process does not occur in the ideal superconductors.

\end{document}